\documentclass[a4paper]{article}

\usepackage[pages=all, color=black, position={current page.south}, placement=bottom, scale=1, opacity=1, vshift=5mm]{background}
\SetBgContents{
	
}      % copyright

\usepackage[margin=1in]{geometry} % full-width

\usepackage{amsmath}
\usepackage{amsthm}
\usepackage{amssymb}
\usepackage{bbold}
\usepackage[utf8]{inputenc}
\usepackage{lmodern}
\usepackage[T1]{fontenc}

\usepackage{hyperref}
\hypersetup{
	unicode,
	pdfauthor={Shayan Roofeh, Vahid Karimipour},
	pdftitle={The Landau-Streater Channel as a Noisy Quantum Channel},
	pdfsubject={The Landau-Streater Channel as a Noisy Channel Model},
	pdfkeywords={article, template, simple},
	pdfproducer={LaTeX},
	pdfcreator={pdflatex}
}

\usepackage[sort&compress,numbers,square]{natbib}
\theoremstyle{plain}

\theoremstyle{definition}
\newtheorem{definition}{Definition}

\usepackage{graphicx}
\graphicspath{{fig/}}
\usepackage{subcaption}
\usepackage{algorithm, algpseudocode} % use algorithm and algorithmicx for typesetting algorithms
\usepackage{mathrsfs} % for \mathscr command
\usepackage{lipsum}
\usepackage{physics}
\def\be{\begin{equation}}
	\def\ee{\end{equation}}
\def\ba{\begin{eqnarray}}
	\def\ea{\end{eqnarray}}
\def\lo{\longrightarrow}
\def\h{\hskip 1cm }

\def\la{\langle}
\def\ra{\rangle}
\def\a{\alpha}

\def\ni{\noindent}

\def\bex{\begin{dinglist}{110}\dsquare}
	\def\eee{\end{dinglist}}
\def\bet{\begin{dinglist}{110}\bsquare}
	\def\bfr{\begin{mdframed}[backgroundcolor=blue!20]\vspace{0.5cm}}
		\def\efr{\vspace{0.5cm}\end{mdframed}}
	
	\title{Construction of channels which in every dimension anti-degrade the depolarizing channel}
	\author{Shayan Roofeh$^1$ \and Vahid Karimipour$^1$\footnote{Corresponding Author: email:vahid@sharif.edu}}
	
	\date{
		$^1$\small{Department of Physics, Sharif University of Technology, Tehran, Iran} \\%
	}
	
\begin{document}
		\maketitle

		%\tableofcontents

		\begin{abstract}
			
			\ni 
			We consider the depolarizing channel in $d$ dimension defined as $D_x(\rho)=(1-x)\rho+x\tr(\rho) \frac{I}{d}$, and explicitly find a quantum channel ${\cal N}_x$ which anti-degrades this, when $x\geq\frac{1}{2}$. This proves that the depolarizing channel $D_x$ has zero capacity when $x\geq\frac{1}{2}$. As a corollary, this implies that any quantum channel when contaminated by white noise stronger than this value loses its capacity completely. 
			Although by arguments based on  symmetric-extendibiliy of the Choi matrix, it is known that the channel is anti-degradable  when $x\geq \frac{d}{2(d+1)}$, \cite{johnson_compatible_2013, Ranade2009Symmetric, Myhr2009Spectrum}, the explicit form of the anti-degrading channel in this larger interval is not known.  We also calculate in closed form the capacity of the complenetary channel ${\cal D}_x^c$ in the region $x\geq \frac{1}{2}$. This adds to the  existing list of quantum channels for which the quantum capacity has been calculated in closed form, see \cite{bennett_capacities_1997},\cite{leditzky_platypus_2023}, and \cite{Roofeh_2024}. 
		\end{abstract}

		%	\noindent\textbf{Keywords:} Werner-Holevo channel, Classical capacity, Quantum Capacity, Flag-Extension.
	
\section{ Introduction}

\ni In the rapidly evolving field of quantum information theory, quantum channels play a fundamental role as the basic mechanisms through which quantum states interact with their environments or are transmitted across different locations. These channels are mathematically described by linear Completely Positive Trace-preserving (CPT) maps. These maps which replace the unitary evolution of closed systems, capture the full range of physical processes that a quantum state may undergo, including the effects of environmental noise, storage of quantum states in quantum memories, general kinds of measurement, and transmission between different places  \cite{wilde_book_2017}. The study of quantum channels is crucial not only for advancing our understanding of quantum mechanics but also for the development of practical applications such as quantum cryptography, quantum communication, and quantum computation \cite{bennett_quantum_2014,ekert_quantum_1991,wilde_book_2017}.
\ni When considered as a model of quantum noise, either in transmission of quantum states through free air or in fibers, or in storage of quantum states in quantum memories, the depolarizing or white noise  \cite{bennett_quantum_2004,holevo_quantum_2020,bennett_capacities_1997,bruss_quantum_2000,king_capacity_2003} stands out as a particularly  important and common yet simple one. In $d$ dimensions it is defined as 
\be
\label{depolarizing}
{\cal D}_x(\rho)=(1-x)\rho + x\tr(\rho)\frac{I_d}{d}
\ee	
where $0\leq x\leq 1$ is the parameter which determines the level of noise. Thus any state is transmitted intact with probability $1-x$ and is turned into a completely mixed state with probability $x$. This type of noise is quite relevant in any experimental platform for quantum computers, where accumulation of any types of noise in individual gates or averaging any type of noise over many directions (i.e. twirling) leads in  large circuits to a simple depolarizing noise in the overall circuit  \cite{Urbanek_2021}. 
 Despite its simplicity, the depolarizing noise encapsulates many of the key challenges in quantum information theory, making it an important channel for study. \\

\ni A central concept in the study of quantum channels is the quantum capacity, which quantifies the maximum rate at which quantum information can be reliably transmitted through a channel. The quantum capacity is a key measure of channel performance  \cite{barnum_information_1998, lloyd_capacity_1997,devetak_private_2005}, directly impacting the efficiency and reliability of quantum communication systems. However, calculating the quantum capacity is notoriously difficult, even for relatively simple channels like the depolarizing channel. This difficulty arises from the generic super-additivity property of most quantum channels \cite{Leditzky_2023} which is the most  challenging problem in calculation of quantum capacities. Unlike classical capacities, where the capacity of two independent channels is simply the sum of their individual capacities, the quantum capacity of a the tensor product of two channels  can be greater than the sum of the capacities of the individual channels.  This  behavior significantly complicates the calculation of quantum capacity because it requires considering not just single uses of the channel, but also the possibility of using entangled input states across multiple channel uses. As a result, the  quantum capacity of a channel 
can only be determined by optimizing the coherent information, over an infinite number of uses of the channel \cite{cubitt_2015}, a task which in general is intractable.  More precisely the quantum capacity of a channel $\Lambda$ should be calculated via \cite{devetak_private_2005}:
\begin{equation}
	\begin{split}
		Q(\Lambda) =
		\lim_{n \to \infty}  \; \frac{1}{n} Q_1(
		\Lambda^{\otimes n}),
	\end{split}     
\end{equation}
where $Q_1(\Lambda)=\max_{\rho} I_c(\rho,\Lambda)$ and the coherent information is given by $I_c(\rho,\Lambda):=S(\Lambda(\rho))-S(\Lambda^c(\rho))$. Here $S(\rho)$ is the von-Neumann entropy $S(\rho):=-\tr \rho \log \rho$. It is known that $Q_1$  is superadditive, i.e. $Q_1(\Lambda_1 \otimes \Lambda_2)\ge Q_1(\Lambda_1)+Q_1(\Lambda_2)$. Therefore the need for the limiting procedure over an infinite number of channel uses, i.e. the so-called regularization, renders an exact calculation of the quantum capacity extremely difficult (or virtually impossible). At the same time this formula provides a lower bound in the form  $Q_1(\Lambda)\leq Q(\Lambda)$ where $Q_1$ is the single shot capacity. \\

\ni We stress that no matter how small the quantum capacity of a channel is, it is still possible to transmit  quantum states through such a channel in an error-free way, albeit with a small rate. However when the capacity vanishes, no matter which encoding we use, transmitting a quantum state through the channel becomes impossible.  Therefore it is important to determine the range of parameters for which a channel has zero quantum capacity. 
An important development in this direction was made by introducing the concepts of degradablity or anti-degradability of quantum channels \cite{devetak_capacity_2005,cubitt_structure_2008}. \\

\begin{figure}[H]
	\centering
	\includegraphics[width=11cm]{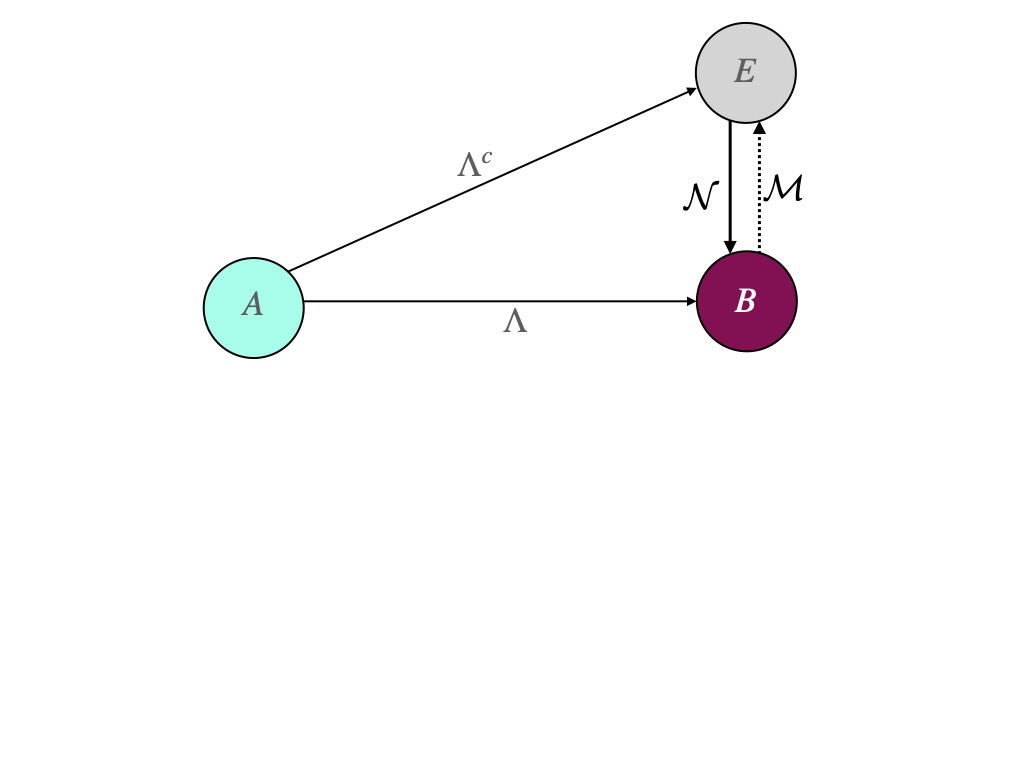}\vspace{-4cm}
	\caption{A channel $\Lambda:A\lo B$ and its complement $\Lambda^c:A\lo E$, where $E$ is the environment. The channel is degradable if ${\cal M}$ exists such that ${\cal M}\circ \Lambda=\Lambda^c$ and is anti-degradable if ${\cal N}$ exists such that ${\cal N}\circ \Lambda^c=\Lambda$. }
	\label{ADFig}
\end{figure}

\begin{definition} \cite{devetak_capacity_2005,cubitt_structure_2008} \label{definitonofdeg} Consider a quantum channel $\mathcal{E}
	\to B$ and its complementary channel $\mathcal{E}^c: A\to E$. The channel $\mathcal{E}$ is said to be degradable if there exists a quantum channel $\mathcal{M}: B \to E$ such that $\mathcal{M} \circ \mathcal{E}(\rho) = \mathcal{E}^c(\rho)$. Conversely, $\mathcal{E}$ is anti-degradable if there is a quantum channel $\mathcal{N}: E \to B$ such that $\mathcal{N} \circ \mathcal{E}^c(\rho) = \mathcal{E}(\rho)$, as shown in figure (\ref{ADFig}). \end{definition}.\\ 

\ni As it is apparent from the definition, in an anti-degradable channel, the environment contains enough information to completely reconstruct the channel's output, which makes reliable transmission of quantum information impossible. This limitation is directly linked to the no-cloning theorem \cite{wootters_single_1982}, which states that it is impossible to create an exact copy of an arbitrary quantum state. If such channels had positive quantum capacity, it would imply a violation of the no-cloning theorem by enabling a process that could perfectly duplicate quantum states.

\noindent On the other hand, if a channel is  degradable, then it possesses the additivity property, and, thus we have  $Q(\Lambda)=Q_1(\Lambda)$ \cite{devetak_capacity_2005}, and in this case the calculation of the quantum capacity becomes a convex optimization problem. \\ 

\ni What we  do in this work is to extend or complement the previous analysis of the quantum depolarizing channel  by explicitly constructing a CPT map that anti-degrades this channel when $x\geq \frac{1}{2}$.  The significance of this type of proof, is that the channel ${\cal D}^c_x$ is then by definition degradable whose capacity is  additive, that is, its one-shot quantum capacity is equal to its full quantum capacity. This allows us to exactly determine the quantum capacity of this one-parameter family of channels. Finally, we can use the fact that concatenation of quantum channels cannot increase the quantum capacity and conclude that the quantum capacity of every quantum channel $\Lambda$, when contaminated by white noise, as in $\Lambda_x (\rho):=(1-x)\Lambda(\rho)+\frac{x}{d}I_d$, drops to zero when $x\geq \frac{1}{2}$.
\\

\noindent The structure of this paper is as follows: In section \ref{pre}, we begin by defining the depolarizing channel and derive its complement channel. Then in section \ref{Ourresults}, we determine  the channel which anti-degrades the depolarizing channel in the region $\frac{1}{2} \le x \le 1$, showing that the capacity of the depolarizing channel is zero in this region.  In the next section we calculate the quantum capacity of the complementary channel in closed form. Finally in section (\ref{arb}) by  using an inequality on quantum channel capacities, we prove that every quantum channel when contaminated by white noise,  loses its capacity, if the white noise is larger than $x=\frac{1}{2}$. The paper is concluded with a discussion.

\subsection{Related Works}
\label{RelatedWorks}

In the absence of exact results, the complexity of the problem has led to a multitude of works in which various techniques have been used to determine lower and upper bounds for quantum capacities of simple channels \cite{kianvash3, poshtvan_capacities_2022, leditzky_quantum_2018}. These include depolarizing and the Gaussian channels  \cite{kianvash1, kianvash2}, the transverse depolarizing channel or the noisy Werner-Holevo channel in three \cite{Roofeh_2024} and in $d$- dimensions \cite{karimipour2024noisylandaustreaterwernerholevochannelarbitrary, Roofeh2025Capacities}.  \\

\ni Of particular interest to us in this work are the Positive Partial Transpose (PPT) channels \cite{horodecki_separability_1997,horodecki_separability_1996}. These channels are characterized by Choi matrices with positive partial transpose and are only capable of producing weakly entangled states that satisfy the PPT condition \cite{peres_separability_1996,smith_quantum_2008,horodecki_separability_1997,horodecki_separability_1996}. This serves as another criterion for identifying the zero quantum capacity region of a channel. In this way one can show that the channel $\mathcal{D}_x$ is PPT if $x \ge \frac{d}{d+1}$. As the dimension gets larger, the PPT region becomes smaller and in very large dimensions it almost vanishes. Therefore this type of argument provides a smaller region of zero-capacity than ours. However, it has been shown in   \cite{myhr_symmetric_2011}, that any channel is anti-degradable if and only if its Choi matrix is symmetric extendible. This idea has been invoked in  \cite{johnson_compatible_2013, Ranade2009Symmetric} to show that the depolarizing channel is anti-degradable and hence has zero capacity in the region   \( \frac{d}{2(d+1)} \le x \leq 1 \), which is larger than the range $x>\frac{1}{2}$, see also \cite{leditzky_useful_2018}). However an explicit construction of a channel which anti-degrades the depolarizing channel, has not been provided up until now.  Finally the anti-degradbility of the channel $D_x$ implies that its complement $D_x^c$ is degradable and hence its one-shot capacity is equal to its full capacity. The full capacity of this new channel $D_x^c$ can now be calculated exactly, as reported in section (\ref{capcom}).  

\section{Depolarizing channel in $d$ dimensions}
\label{pre} 
The Kraus representation of the depolarizing channel (\ref{depolarizing}) is given by 
\be
{\cal D}_x(\rho)=A_0\rho A_0^\dagger + \sum_{i,j=1}^dA_{ij}\rho A_{ij}^\dagger
\ee	
where
\be
A_0=\sqrt{1-x}I_d, \ \ A_{ij}=\sqrt{\frac{x}{d}}|i\ra\la j|.
\label{KrausDep}
\ee
Here $\{|i\ra, i=1,\cdots d\}$ are the orthonormal computational basis states of the $d$ dimensional Hilbert space $H_d$ and $|0\ra$ is an extra normalized state, independent from the states $|i\ra$ for all $i$.  \\

\ni The notion of the complement of a quantum channel is based on the well-known Stinespring dilation theorem \cite{stinespring}. This theorem states that a quantum channel $\Phi: A \to B$ can be realized as a unitary map $U: A \otimes E \to B \otimes E'$, where $E$ and $E'$ represent the environments associated with $A$ and $B$, respectively. Explicitly, this can be expressed as:
 \begin{equation} \Phi(\rho) = \text{tr}_{E'}(U\rho U^{\dagger}), \end{equation} 
  where $U$ is an isometric map from $A$ to ${B} \otimes E'$. Under this formalism, the complementary channel $\Phi^c: A \to E'$ is defined by: 
 \begin{equation} \label{steincomp} \Phi^c(\rho) = \text{tr}_{B}(U\rho U^{\dagger}), \end{equation}
   which describes a map from the input system to the environment of the output system (see figure (\ref{ADFig})). It is crucial to emphasize that the complement of a quantum channel is not uniquely defined, though different complements are related via isometries, as discussed in \cite{datta_complementarity_2006}. The complementary channel of an arbitrary channel $\Phi:A \to B$ with Kraus operators $\{K_i\}$ can be obtained by the following isometric map:
$$U\ket{\psi}=\sum_i K_i \ket{\psi}\otimes \ket{i}$$
Then the the complementary channel of this channel can be determined as:
\begin{equation}
\Phi^c(\rho)=\sum_{i,j} \Tr (K_i \rho K_j^\dagger) \dyad{i}{j}
\end{equation}
To find the complement of the depolarizing channel \cite{datta_complementarity_2006, smaczynski2016selfcomplementary}, we use the isometric extension ${\cal E}_{A\lo BE}$ which is defined as
\ba
{\cal E}_{A\lo BE}(Y)&=&A_0 Y A_0^\dagger \otimes |0\ra\la 0|+ \sum_{i,j=1}^d A_0 Y A_{ij}^\dagger \otimes |0\ra\la i,j|\cr
&+&  \sum_{i,j=1}^d A_{ij} Y A_{0}^\dagger \otimes |i,j\ra\la 0|+ \sum_{i,j,r,s=1}^d A_{ij} Y A_{rs}^\dagger \otimes |i,j\ra\la r,s|,
\ea
where the Hilbert space $H_E$ is $1+d^2$ dimensional, whose elements are the {\it formal} linear combinations of the orthonormal states $\{|0\ra, |i,j\ra, \ i,j=1\cdots d\}.$ Thus, \be \la 0|i,j\ra=0,\ \  \la i,j|k,l\ra=\delta_{i,k}\delta_{j,l}.\ee  By taking the trace over the system $B$ and using the explicit form of the Kraus operators (\ref{KrausDep}), we obtain the complementary channel ${\cal D}^c: A\lo E$. Thus by noting that  
\be
\tr(A_0YA_0^\dagger)=(1-x)\tr(Y),\ \ \ \ \  \tr(A_0YA_{ij}^\dagger)=\sqrt{\frac{x(1-x)}{d}}Y_{ji}, \ \ \ \    \tr(A_{ij}YA_{rs}^\dagger)=\frac{x}{d}Y_{jr}\delta_{is} 
\ee
we obtain 
\be\label{dxc}
{\cal D}_x^c(Y)=(1-x)\tr(Y)|0\ra\la 0|+\xi \sum_{i,j}\big(Y_{ij}|0\ra \la i,j|+Y_{ji}|i,j\ra\la 0|\big)+\frac{x}{d}\sum_{i,j,k}Y_{jk}|i,j\ra\la i,k|,
\ee 
where $\xi=\sqrt{\frac{x(1-x)}{d}}$ and $|Y\ra$ is the vectorized form of the matrix \(Y\), i.e. $\ket{Y}=\sum_{i,j} Y_{ij} \ket{ij}$
This can be rewritten in matrix form as:
\be\label{Dxc}
{\mathcal{D}}_x^c(Y) =\left( 
\begin{array}{c|c}
	(1 - x) \operatorname{Tr}(Y) & \xi \bra{Y}^* \, \\ \hline
	\xi\, \ket{Y^T} & \frac{x}{d}(I_d \otimes Y)
\end{array}
\right)
\ee
\ni or more compactly as 
\be
{\cal D}_x^c(Y)=(1-x)\tr(Y)|0\ra\la 0|+\xi \big(|0\ra \la Y|^*+|Y^T\ra\la 0|\big)+\frac{x}{d}I_d\otimes Y.
\ee
In the next section we show that this depolarizing channel $D_x$ is anti-degradable in the range $x\in [\frac{1}{2},1]$ which implies that the channel $D_x^c$ is degradable in this range. This means that its full quantum capacity whose calculation is a formidable task, is equal to its one-shot quantum capacity.  In section (\ref{capcom}),  we will determine this capacity in closed form. 

\section{The  channel which anti-degrades the depolarizing channel}
\label{Ourresults}
\ni In this section we explicitly find the form of the channel that anti-degrades the depolarizing channel. We present our results in the form of a theorem and its corollary. \\

\ni {\bf Theorem:} The depolarizing channel \(\mathcal{D}_x\) is anti-degradable and therefore has zero capacity for \(x \geq \frac{1}{2}\).\\

\ni {\bf Proof:}
The channel ${\cal D}_x$ is anti-degradable if we can find a quantum channel, or completely trace-preserving map ${\cal N}: E\lo B$, such that the following equality holds \cite{devetak_capacity_2005}
\be\label{ADN}
{\cal N}\circ {\cal D}_x^c(Y)={\cal D}_x(Y).
\ee
We define the channel ${\cal N}$ by its Kraus representation,
\be\label{NR}
{\cal N}(Z)=\sum_{\a}R_\a Z R_\a^\dagger, \h Z\in M_{d^2+1}
\ee
where $M_{d^2+1}$ is the space of $d^2+1$ matrices and the Kraus operators $R_\a\in \{M_k, N_k, Q_{i,j,k}\}$ are taken to be as 

\ba\label{MNQ}
M_k&=& \frac{1}{\sqrt{d}} |k\ra \la 0|\cr
N_k&=&\beta\sum_{j}|j\ra\la k,j|\cr
Q_{i,jk}&=&\delta |i\ra\la j,k|.
\label{KrausN}
\ea
Here $\beta$ and $\delta$ are real parameters which are to be determined so that the condition of equation (\ref{ADN}) is satisfied.
Note that these parameters can be taken to be real, since any complex phase will be cancelled out in the final action of the channel (\ref{NR}). 
We can then show that the property (\ref{ADN}) holds, provided that the following inequality is true
\be
d\delta^2=\frac{2x-1}{x}.
\ee
In view of the reality of $\delta$, this shows that if $x\geq  \frac{1}{2}$, we can find the channel ${\cal N}$ such that ${\cal N}\circ {\cal D}_x^c={\cal D}_x$, hence the channel is anti-degradable and has zero quantum capacity. \\

\ni We first check the trace-preserving property, namely
\be
\sum_\a R_\a^\dagger R_\a\equiv \sum_{k}(M_k^\dagger M_k + N_k^\dagger N_k)+\sum_{i,j,k}Q_{i,jk}^\dagger Q_{i,jk}=|0\ra\la 0| +(\beta^2+d\delta^2)\sum_{j,k}|j,k\ra\la j,k|=I_{E},
\ee
which imposes the following condition on the parameters
\be
\beta^2+d\delta^2=1.
\ee
To find the combined channel ${\cal N}\circ {\cal D}_x$, we calculate the effect of each class of Kraus operators on ${\cal D}_x^c(Y)$ separately. It is readily found that 

\be\label{M}
\sum_k M_k{\cal D}_x^c(Y)M_k^\dagger = \frac{1}{d}(1-x)\tr(Y)|k\ra\la k|=\frac{1-x}{d}\tr(Y) I_d
\ee
It is also straightforward to show that 
\be\label{N1}
\sum_k N_k{\cal D}_x^c(Y)N_k^\dagger = \beta^2 xY.
\ee
To see this, we note that 
\be\label{NN}
\sum_k N_k{\cal D}_x^c(Y)N_k^\dagger=\beta^2 \sum_k \sum_{j,j'}|j\ra\la k,j|{\cal D}_x^c(Y)|k,j'\ra \la j'|
\ee
We need to calculate the middle matrix element which is 
\ba\label{Matrix1}
&&\la k,j|{\cal D}_x^c(Y)|k,j'\ra=\sum_{m,n,p}\la k,j|\frac{x}{d}Y_{n,p}|m,n\ra\la m,p|k,j'\ra\cr
&=& \sum_{m,n,p}\frac{x}{d}Y_{np}\delta_{k,m}\delta_{n,j}\delta_{m,k}\delta_{p,j'}=\frac{x}{d}\sum_{m}Y_{j,j'}\delta_{k,m}\delta_{m,k}=\frac{x}{d}Y_{j,j'}.
\ea
Inserting this matrix element in (\ref{NN}), we arrive at (\ref{N1}). Finally, for the third class of Kraus operators, we find 

\be\label{Q}
\sum_{i,j,k} Q_{i,jk}{\cal D}_x^c(Y)Q_{i,jk}^\dagger = \delta^2 x \tr(Y)I_d.
\ee
To show this we note that 
\be\label{Q1} 
\sum_{i,j,k} Q_{i,jk}{\cal D}_x^c(Y)Q_{i,jk}^\dagger=\sum_{i,j,k}|i\ra\la j,k|{\cal D}_x^c|j,k\ra\la i|
\ee
for which we need the following matrix element which from (\ref{Matrix1}) is given by
\be
\la j,k|D_x^c(Y)|j,k\ra=\frac{x}{d}Y_{k,k}.
\ee
Inserting this in (\ref{Q1}) leads to (\ref{Q}). Summing the effects of all three classes of Kraus operators, namely (\ref{M}, \ref{N1}, and \ref{Q}), we come to 

\be
{\cal N}\circ {\cal D}_x^c(Y)=\left[\frac{1-x}{d}+\delta^2 x \right]\tr(Y)I_d+\beta^2xY
\ee
Equating this with the channel $D_x(Y)=(1-x)Y + \frac{x}{d}\tr(Y)\frac{I_d}{d}$ and observing the trace-preserving condition, we should have:
\ba
&&\frac{1-x}{d}+\delta^2 x=\frac{x}{d}\cr
&&\beta^2x=(1-x)\cr
&&\beta^2+d\delta^2=1.
\ea
Note that we have three equations for the two unknowns $\beta$ and $\delta$, but the equations are not independent. Inserting the values of $\beta$ in the other two equations, they both lead to 
\be
d\delta^2=\frac{2x-1}{x},
\ee
This proves that if $x\geq \frac{1}{2}$, the map ${\cal N}$ exists and hence the channel ${\cal D}_x$ is anti-degradable, which was to be proved.\\

\ni It is instructive to find how the  channel ${\cal N}$, acts on arbitrary $d^2+1$ dimensional states. Let $$Z=\begin{pmatrix} z_0& {\bf z}\\ {\bf z}^\dagger& {\cal Z} \end{pmatrix}$$ be such a state or more generally an arbitrary matrix in $M_{d^2+1}$.  Then from the explicit form of the Kraus operators in (\ref{KrausN}), we find
\be
{\cal N}(Z)=\frac{1}{d}\sum_{k}|k\ra\la 0|Z|0\ra\la k|+\beta^2 \sum_{j,k,l}|j\ra\la k,j|Z|k,l\ra\la l|+\delta^2\sum_{i,j,k}| i\ra\la k,j|Z|j,k\ra\la i|.
\ee
Simplifying this, and inserting the values of $\beta$ and $\delta$, we obtain
\be
\label{DegradingChannel}
{\cal N}(Z)=\frac{1}{d}z_0 I_{d}+ \frac{1-x}{x} \tr_1 {\cal Z}+\frac{2x-1}{dx} \tr({\cal Z})I_{d}
\ee
where $I_d=\sum_k|k\ra\la k|$ and $\tr_1$ denotes the partial trace on the first factor.  It is now quite obvious why  this channel anti-degrades the depolarizing channel when $x\geq \frac{1}{2}$. Furthermore, we now have a new family of channels, namely $\mathcal{D}_x^c$, which are all degradable and hence enjoy additivity; that is, their quantum capacity is equal to their one-shot capacity.

\section{The Quantum Capacity of the Complementary Channel}\label{capcom}

\ni In the region \( \frac{1}{2} \le x \le 1 \), the complementary channel is degradable, which ensures that its coherent information is both additive and concave \cite{cubitt_structure_2008,Yard2008Capacity}. Specifically, we have
\[
I_c(\mathcal{D}_x^c, \sum_i p_i \rho_i) \ge \sum_i p_i I_c(\mathcal{D}_x^c, \rho_i),
\]
where \( I_c \) denotes the coherent information, \( \mathcal{D}_x^c \) is the complementary channel, and \( \rho_i \) are the input states with probabilities \( p_i \).

\begin{figure}[H]
 	\centering
 	\includegraphics[width=13cm]{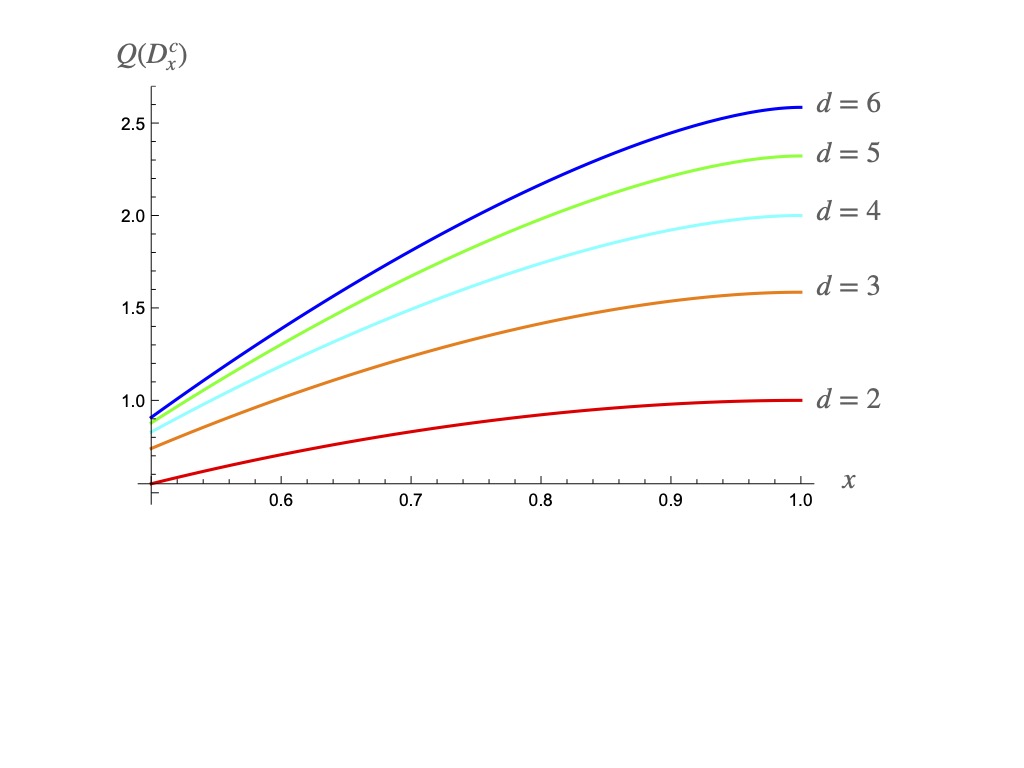}\vspace{-3cm}
 	\caption{The exact quantum capacities of the channel $D_x^c$ as a function of $x\in [1/2,1]$ in different dimensions.  In this region, the quantum capacity of the depolarizing channel $D_x$ is exactly zero. }
 	\label{plots}
 \end{figure}

\ni To find the coherent information $I_c$, we note that the depolarizing channel itself is covariant under the unitary group $U(d)$, that is 
$$D_x(U_g\rho U_g^\dagger)=U_gD_x(\rho)U_g^\dagger\h \forall g\in U(d),$$
where $U_g$ is the defining representation of the group element $g$ in the unitary group. 
In \cite{poshtvan_capacities_2022} it has been proved that the covariance of every channel induces the covariance, albeit with different representations of the same group on the complementary channel, that is 
$$D^c_x(U_g\rho U_g^\dagger)=\Omega_gD^c_x(\rho)\Omega_g^\dagger \h \forall g\in U(d),$$
where $\Omega(g)$ is now a different representation. 
  Given that the representation $U_g$ is still an irreducible representation. and by  applying Schur's lemma, we find:
\[
I_c(\mathcal{D}_x^c, \rho_0) = \int_{g \in U(d)} dg  I_c(\mathcal{D}_x^c, U_g \rho_0 U_g^\dagger) \le I_c\left(\mathcal{D}_x^c, \int_{g \in U(d)} dg  U_g \rho_0 U_g^\dagger\right) = I_c\left(\mathcal{D}_x^c, \frac{I}{d}\right),
\]
where $\int_{g\in U(d)} dg$  is the Haar measure  and $\frac{I}{d}$ is the maximally mixed state. Now we should determine eigenvalues of $\mathcal{D}_x^c(\frac{I}{d})$:
\[
{\mathcal{D}}_x^c(\frac{I}{d}) =\left( 
\begin{array}{c|c}
	(1 - x) & \frac{\xi}{d} \bra{\phi^+} \, \\ \hline
	\frac{\xi}{d}\, \ket{\phi^+} & \frac{x}{d}(I_d \otimes I_d)
\end{array}
\right),
\]
where $\ket{\phi^+}=\sum_i \ket{ii}$. The eigenvalues of this matrix are as following:\\
\be 	\text{Spectrum}(\mathcal{D}_x^c(\frac{I}{d}))=
\begin{cases}
		0,  &\ \ \ \   -\frac{\xi}{1-x}\ket{0}+\ket{\phi^+} \\
        \frac{x}{d^2}+(1-x), &\ \ \ \   \frac{d(1-x)}{\xi}\ket{0}+\ket{\phi^+} \\
	\frac{x}{d^2} & \ \ \ \ \ket{ij} \ \ \ \h \ \ i\neq j \\
	\frac{x}{d^2},  &\ \ \ \   |i,i\ra-|i+1,i+1\ra\ \ \ \ i=1 \cdots d-1.
\end{cases}
\ee 
Consequently, we can write:
$$S(\mathcal{D}_x^c(\frac{I}{d}))=-(d^2-1)\frac{x}{d^2}\log(\frac{x}{d^2})-(\frac{x}{d^2}+1-x)\log \left(\frac{x}{d^2}+1-x \right)$$
Hence:
\begin{equation}
Q(\mathcal{D}_x^c)=I_c(\mathcal{D}_x^c, \frac{I}{d}))=-(d^2-1)\frac{x}{d^2}\log(\frac{x}{d^2})-(\frac{x}{d^2}+1-x)\log \left(\frac{x}{d^2}+1-x \right)-\log(d)
\end{equation}
This expression is valid only in the range $x\in [\frac{1}{2},1]$, where the channel $D_x^c$ is degradable and its one-shot capacity is equal to its full quantum capacity. The interval $x\in[0,\frac{1}{2}]$ only gives a lower bound for the full quantum capacity. Note that this is an increasing function of $x$ and at the special point $x=1$, it reduces to 
\be
Q(\mathcal{D}_1^c)=\log d,
\ee 
as expected. The reason is that at this point the depolarizing channel converts any input state into a maximally mixed state and all the information goes to the environment. This is evident from equation (\ref{dxc}), which shows that at $$D_1(\rho)=\frac{1}{d}I_d,\h D_1^c(\rho)=\frac{1}{d}I_d\otimes\rho.$$ In fact at this point the complementary channel completely preserves the input state, while the channel itself completely destroys it. On the contrary, at $x=0$, the channel preserves the state while the complementary channel destroys it, i.e. 
$$D_0(\rho)=\rho,\h D_0^c(\rho)=\tr(\rho)|0\ra\la 0|.$$ 
 Therefore at this point the quantum capacity of the channel is maximum while that of the complementary channel vanishes.  However this point is not in the region $x\in [1/2,1]$, where the channel $D_x^c$ is degradable. Outside this interval, the one-shot quantum capacity gives only a lower bound for the full capacity.  Figure (\ref{plots}) shows the quantum capacity of $D^c_x$ as a function of $x$ in several dimensions.
 
\section{Discussion}
\label{discussion}
In this paper, we have found the explicit form of a channel which anti-degrades the depolarizing channel (\ref{depolarizing} ) in the range $x\in [\frac{1}{2},1]$.  This already shows that when $x\geq \frac{1}{2}$, the  quantum capacity of this channel is zero. Moreover, this shows that the complementary channel $D_c^c$ given in (\ref{Dxc}) is degradable and hence its quantum capacity can be calculated in closed form.  
 Calculation of quantum capacity of this channel does not require any regularization and is equal to their one-shot capacity.  In addition, the anti-degradability of the depolarizing channel can be used to show that the quantum capacity of every quantum channel, which is contaminated by white (depolarizing) noise vanishes, if the strength of the noise is large enough. In fact given any arbitrary channel $\Lambda$, and knowing that  the quantum capacity of channels never increases by concatenating them \cite{schumacher_quantum_1996}, we  obviously have the following inequality on quantum capacity of channels, namely \\
\be\label{data}
Q( {\cal D}_x \circ \Lambda)\leq Q({\cal D}_x).
\ee
Since the channel $\Lambda$ is trace-preserving, the quantum channel $\Lambda_x:={\cal D}_x\circ \Lambda$ acts as follows:
\be
\Lambda_x(\rho)=(1-x)\Lambda(\rho)+{x}\tr(\rho)\frac{I_d}{d}
\ee 
which when combined with the previous inequality leads to 
\be
Q(\Lambda_x)\equiv Q({\cal D}_x\circ \Lambda)=0\h {\rm if}\ x\geq \frac{1}{2}.
\ee
Additionally, we have shown  that every quantum channel $\Lambda$ and in every dimension, when contaminated by depolarizing noise in the form  \(\Lambda_x(\rho)=(1-x)\Lambda(\rho) + \frac{x}{d}I_d \text{Tr}(\rho)\) also has zero capacity for \(\frac{1}{2} \le x \le 1\). This provides a large family of zero-capacity quantum channels in $d$ dimensions which can motivate further study regarding the interesting problem of superactivation of quantum capacities \cite{Smith_2008, Leditzky_2023, Yard2008Capacity}.  This is a curious property of entanglement \cite{horodecki_bound_1999} and quantum information processing, where two zero-capacity channels ${\cal E}_1$ and ${\cal E}_2$ when used together in the form ${\cal E}_1\otimes {\cal E}_2$ can have non-zero capacity. There are few examples of channels which show this property \cite{Smith_2008}. There are even specifically constructed zero capacity qutrit channels, called Platypus channels in  \cite{Leditzky_2023,Leditzky_2023a}, which can activate many other types of qubit channels, including amplitude damping, erasure, depolarizing, and random qubit channel. Besides these $d-$ dimensional channels, we have shown in a recent work \cite{Roofeh_2024} that a one-parameter extension of the qutrit Werner-Holevo channel or the transpose-depolarizing channel of the form
$${\cal E}_x(\rho)=(1-x)\rho + \frac{x}{2}(\tr(\rho)I-\rho^T)$$
are anti-degradable and thus has zero quantum capacity in the range $\frac{4}{7}\leq x\leq 1$. 
 Invoking the inequality (\ref{data}) again, we find a large family of zero-capacity channels in the form
$$\Phi_x(\rho)=(1-x)\Phi(\rho) + \frac{x}{2}(\tr(\rho)I-\Phi(\rho)^T),$$ where $\Phi$ is any CPT map. It would be then interesting to 
explore the problem of super-activation with these two types of channel at hand. \\
 
  Finally we remark about the relation of our work with the works of \cite{johnson_compatible_2013, Ranade2009Symmetric, Myhr2009Spectrum} . 
 Although the $d$- dimensional depolarizing channel is known to be anti-degradable in the parameter region $ x\in\left[\frac{d}{2(d+1)},1\right]$, the explicit form of an anti-degrading map is, to the best of our knowledge, not available in closed form for arbitrary dimension. Existing proofs of anti-degradability rely mainly on symmetry and covariance arguments, together with positivity of the associated Choi operator, which establish the existence of a completely positive trace-preserving map without yielding a simple Kraus representation. Extracting explicit Kraus operators would in general require diagonalization of highly structured $d^{2}\times d^{2}$ Choi matrices, a task that rapidly becomes algebraically cumbersome as the dimension increases. Instead, we have proceeded by a judicious guesswork and have constructed the Kraus representation of an anti-degrading channel in eqs. (\ref{MNQ}) from the very beginning. The price is that this channel anti-degrades the channel $D_x$ in the smaller region $x\in [\frac{1}{2},1]$. We have indeed tried many  generalizion of the channel (\ref{MNQ}) in order to extend this region, and curiously enough, all these generalizations lead to the same interval. It therefore remains an interesting and challenging problem to find the explicit form of the anti-degrading channel in the region $x\in [\frac{d}{2(d+1)},1] $. 
 Consequently, the anti-degrading map is presently understood more at an abstract structural level than in a constructive form.

\section*{Funding}

This research was supported by the Iran National Science Foundation under Grant No. 4022322.

\section*{Data Availability}

All data present within the manuscript.

	\bibliography{refs}
	\pagebreak

\end{document}